\documentclass[9pt,twocolumn,twoside,lineno]{article}

\newcommand{\IdC}{\text{IdC}}
\newcommand{\vmax}{v_+}
\newcommand{\vmin}{v_-}
\newcommand{\umax}{u_+}
\newcommand{\umin}{u_-}
\usepackage{xfrac}
\usepackage[load-configurations = abbreviations]{siunitx}

\usepackage{amsmath} 
\usepackage[english]{babel} 
\usepackage[hang, small,labelfont=bf,up,textfont=it,up]{caption}\usepackage{booktabs} 
\usepackage{lettrine} 
\usepackage{enumitem} 
\setlist[itemize]{noitemsep} 

\usepackage{abstract} 
  
\usepackage{titlesec} 
\renewcommand\thesection{\Roman{section}} 
\renewcommand\thesubsection{\roman{subsection}} \titleformat{\section}[block]{\large\scshape\centering}{\thesection.}{1em}{} \titleformat{\subsection}[block]{\large}{\thesubsection.}{1em}{} 
\usepackage{fancyhdr} 
\usepackage{titling} 

\usepackage{hyperref} 

\pretitle{\begin{center}\Huge\bfseries} 
\posttitle{\end{center}} 
\title{Gait transition in swimming}
\author{%
\textsc{Remi Carmigniani\thanks{Corresponding author}} \\[1ex] 
\normalsize Ecole des Ponts ParisTech \\ 
\normalsize \href{mailto:remi.carmigniani@enpc.fr}{remi.carmigniani@enpc.fr} 
\and 
\textsc{Ludovic Seifert \& Didier Chollet} \\[1ex] 
\normalsize CETAPS EA3832, Faculty of Sports Sciences, University of Rouen Normandy
\and 
\textsc{Christophe Clanet} \\[1ex] 
\normalsize LadhyX, Ecole Polytechnique
}

\date{\today} 
\renewcommand{\maketitlehookd}{%
\begin{abstract}
The skill to swim fast results from the interplay between generating high thrust while minimizing  drag. 
In front crawl, swimmers achieve this goal by adapting their inter-arm coordination according to the race pace. A transition has been observed from a catch-up pattern of coordination (i.e. lag time between the propulsion of the two arms) to a superposition pattern of coordination as the velocity increases. 
Expert swimmers choose a catch-up coordination pattern at low velocities with a constant relative lag time of glide during the cycle and switch to a maximum propulsion force strategy at higher velocities. 
This transition is explained using a burst-and-coast model. At low velocities, the choice of coordination can be understood through two parameters: the time of propulsion and the gliding effectiveness. These parameters can characterize a swimmer and help to optimize their technique.
\end{abstract}
}

\begin{document}

\maketitle

\begin{figure}[h]
\centering
\includegraphics[width=0.8\linewidth]{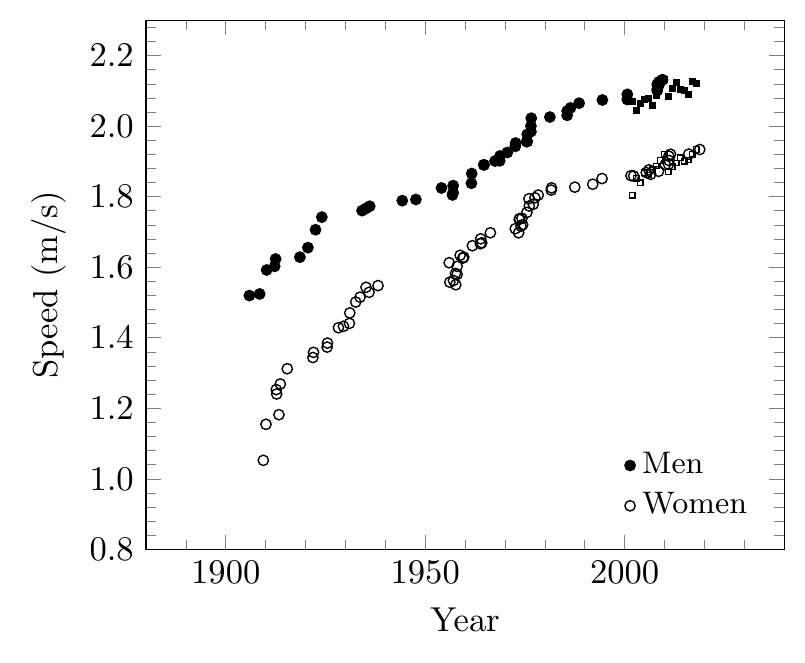}
\caption{Evolution of the mean velocity over time of the 100 m long course freestyle. The circles denote the world records evolution. The squares denote the year best performance from 2001 to today.}
\label{evolSpeed100mLC}
\end{figure}

{A}lthough we can find evidences of swimming in the artwork of ancien Egypt over 2,000 BC, modern competitive swimming started in the early 19th-century England \cite{McVicar1936}. The search for speed in swimming led to changes of the technique from the natural quadrupeds dog fashion technique to the breaststroke, then side-stroke and Trudgen-stroke, all the way to the modern front-crawl. The front-crawl was pioneered in competition by the Australian Richard Cavill at the beginning of the 20th century. He was largely inspired by natives surfers from the Solomon Islands \cite{McVicar1936}. The technique was refined over time as the average speed of swimmers has continued to increase over the century (see figure \ref{evolSpeed100mLC}). Front-crawl is now used on a large range of distances in swimming pool and open-water races. It appears to be the most efficient swimming technique as it is the only one used for long distances (over 200 m) and the fastest one (used in freestyle sprint)\cite{BARBOSA2010}. It is characterized by alternated arm propulsion phases and arm recovery out of the water.

The skill to swim fast is a combination between generating high thrust and minimizing drag due to aquatic resistance on the body. The first study investigating drag during human locomotion in water can be traced back all the way to the early beginning of the 20$^\text{th}$ century \cite{katzenstein1905arch}. Karpovich \cite{karpovich1933water} pioneered the quantification of human body drag using a towing protocol (called passive drag, $D_{p,b}$). He found that the passive drag when the swimmer is fully extended in a so called streamline position near the surface was about $D_{p,b} = k_{p,b} v^2$, where $v$ denotes the towing velocity and $k_{p,b} \approx 31$ (24) \si[per-mode=symbol]{\kilogram\per\metre} for men (women, respectively). Then numerous research examining passive drag have emerged as shown in the review of Scurati \textit{et al.} \cite{scurati2018}. The mean drag experienced during swimming is still not fully understood and continue being investigated\cite{di1974energetics,kolmogorov1992,toussaint1988active,narita2017developing}. A simple way to reduce the drag is to swim in the wake of another swimmer \cite{Westerweel2016,silva2008analysis,chatard2003drafting}. This is called drafting and the effects of drafting on the swimmer technique and race strategy are still to explore.

The swimming performance is solely evaluated on the time to reach a certain distance. To understand the link between the achieved performance and the swimming technique, researchers have first focused their attention on the arm stroke frequency (also called stroke rate) $f_R$, and the mean velocity of the swimmer $\overline{v}$ \cite{craig1985velocity,craig1979relationships}. To link these two quantities, they defined the distance per stroke (or stroke length) $L_s = \overline{v}/f_R$. Craig \& Pendergast \cite{craig1979relationships} collected data on expert swimmers asking them to swim at a given velocity using the minimum stroke frequency they could achieve. They observed that swimmers did not use this minimum stroke rate technique for long distance races (over 200 m). They commented that even though these swimmers could achieve the same velocity with a lower frequency (and hence a longer stroke length), they used a higher frequency and a lower force per stroke to reduce fatigue. 
Costill \textit{et al.} \cite{costill1985energy} emphasize the importance of stroke technique on the performance and defined a stroke index $SI = \overline{v} L_s$ to evaluate the swimming economy. 

\begin{figure*}[h]
\centering
\includegraphics[width=15cm]{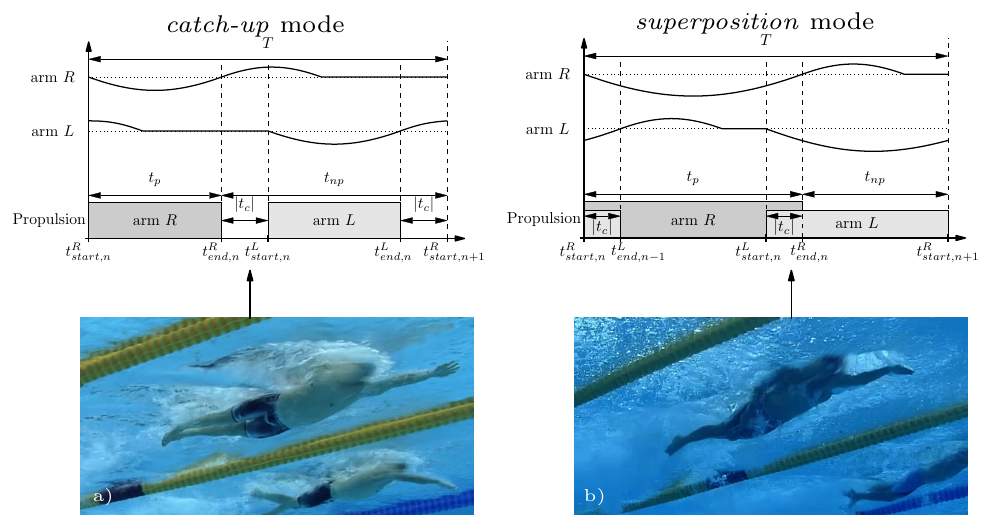}
\caption{Main differences between long (left) and short (right) distance swimmers' coordination patterns. Photos are extracted from races at the Olympic Games with the permission of The Olympic Multimedia Library.}
\label{diffLongSprint}
\end{figure*}

Focusing furthermore on the swimming technique, Chollet \textit{et al.} \cite{chollet2000} investigated the arm stroke phase organization during a stroke cycle and defined the index of coordination ($\IdC$). This non-dimensional number characterizes the temporal motor organization of propulsion phases. 
The two main patterns of coordination can be simplified to the sketches of figure \ref{diffLongSprint}. The solid lines represent a simplified hand elevation compared to the mean water level (dotted lines)\footnote{Note that when the solid curve overlaps the dotted one, this intends to mean that the hand has entered the water but is not yet active in the propulsion.}. When the solid curve is below this level, a propulsive phase occurs. This is further outlined by the gray blocks at the bottom. The arms are identified by the index $i\in\left\{L,R\right\}$, for left and right, respectively. This index enables to track the successions of propulsive phases. 
As an example, we consider the nth cycle of the right arm. It begins at $t_{start,n}^R$ and ends when this arm starts its next propulsive phase $t_{start,n+1}^{R}$. The cycles repeat periodically with a period $T = t_{start,n}^{R}-t_{start,n}^{R}$. The propulsion phase of one arm lasts $t_p= t^R_{end,n} - t^R_{start,n}$ and the non-propulsive phase $t_{np}= t^{R}_{start,n+1} - t^R_{end,n}$. The coordination time is then defined by:
\begin{align}
t_c = t^{R}_{end,n}-t^{L}_{start,n},
\end{align}
and the index of coordination corresponds to the non-dimensional time of coordination compared to the cycle period:
\begin{align}
\IdC = t_c/T.
\end{align}
In the case of figure \ref{diffLongSprint}-a), \textit{catch-up} mode, the index of coordination is negative as the propulsive phase of the latter arm starts after the end of the propulsive phase of the former. This technique is exhibited by long distance swimmers who used glide within the cycle. During this glide they adopt a streamline arm position as illustrated in the picture of figure \ref{diffLongSprint}-a). On the contrary, in figure \ref{diffLongSprint}-b), \textit{superposition} mode, the index of coordination is positive. There is a time $2\left|t_c\right|$ during which both arms performed their propulsion.
A third pattern of coordination can be defined at the transition between the former two and is referred to \textit{opposition} in the literature.
It corresponds to the case where one arm starts its propulsion phase when the other finishes. There is no time lag between the two propulsion phases ($\IdC = 0$).
These three distinctive patterns of coordination were first described by Costill \textit{et al.} \cite{trove.nla.gov.au/work/7341874} and then quantified by Chollet \textit{et al.} \cite{chollet2000}. They observed the choice of coordination of different level swimmers. Expert swimmers were able to reach higher swimming velocity thanks to higher positive index of coordination than non-expert swimmers both on incremental tests \cite{chollet2000,seifert2007swimming} and 100-m races \cite{seifert2007kinematic}. The effect of fatigue on the coordination was also investigated by Alberty \textit{et al.} \cite{alberty2009stroking}. 
They observed a general increase of the index of coordination with fatigue.
A physical model discussing the motor coordination is proposed in our current study to understand the transition from \textit{catch-up} to \textit{superposition} mode and the optimal choice of coordination depending on the targeted velocity of swimming.

Our study is organized in two steps. First, we present the field observations of expert swimmers coordination and discussed a simple way to compare the swimmers among them. The swimmers used only their arms to generate thrust. As previously noticed, for low velocity, hence long distance races pace, the swimmers prefer a catch-up mode of swimming. 
Second, we propose a physical model to understand this choice of coordination. 
The model is compared to our field observations and a linearized expression is derived.

\section*{Field investigation and first coordination model}
\subsection*{Raw observations}
Following the work of Chollet \textit{et al.} \cite{chollet2000}, we consider the motor coordination of national level French swimmers. To simplify the discussion, the motor coordination of the swimmers is averaged between the two arms and the legs motions are ignored. That is to say that the swimmers are considered symmetrical and only the arms coordination are discussed. 

In the current study, we consider the data collected on 16 French male swimmers in 2007 for whom the mean $\pm$ standard deviation (min, max) of age, body mass, height, arm span and arm length were: 21.2$\pm$4.4 (19, 31) years, 78.8$\pm$ 8.5 (66.3,90.5) \si[per-mode=symbol]{\kilogram}, 1.84$\pm$0.03 (1.70, 1.93) \si[per-mode=symbol]{\metre}, 1.91$\pm$0.08 (1.70, 2.14) \si[per-mode=symbol]{\metre}, 0.65$\pm$0.05 (0.60, 0.75) \si[per-mode=symbol]{\metre}, respectively. At the time of the experiment, they were practicing a minimum of 10 hours a week and had been swimming competitively for 12.1$\pm$3.5 years, confirming their expert level \cite{ericsson1996expert}. An extra swimmer was also tested. He was highly specialised in sprint race (50 m race). He had performed similar coordination test in the past and showed behaviour similar to the one described in the paper. For this test, he surprisingly performed drastically differently than before. We decided to remove him from the data set due to this change of coordination, which was probably due to specific sprint work.

The personal best record of the 16 expert swimmers was in average 54.2 $\pm$1.8 (50.33, 57.8) \si[per-mode=symbol]{\second} at the 100-m freestyle in the long pool. All swimmers competed at national level. They were all tested on two graded speed tests in a randomized order using only the arms in front crawl. Their legs were tied and they were equipped with a pull buoy for buoyancy. They all volunteered for this study and gave their written consent to participate. 

One test consists of simulated racing techniques where these expert swimmers were asked to swim at 8 different velocities corresponding to different race paces (from 3000 m to 50 m + maximal speed) on a single 25 m lap. During this test, the swimmers were video recorded by two synchronized underwater video cameras at 50 fps (Sony compact FCB-EX10L), in order to get a front and side view, from which the different stroke phases and the arm coordination have been computed. The protocol is similar to the one described in \cite{chollet2000, seifert2007,seifert2010}.

\begin{figure}[h!]
\centering
\includegraphics[width=.8\linewidth]{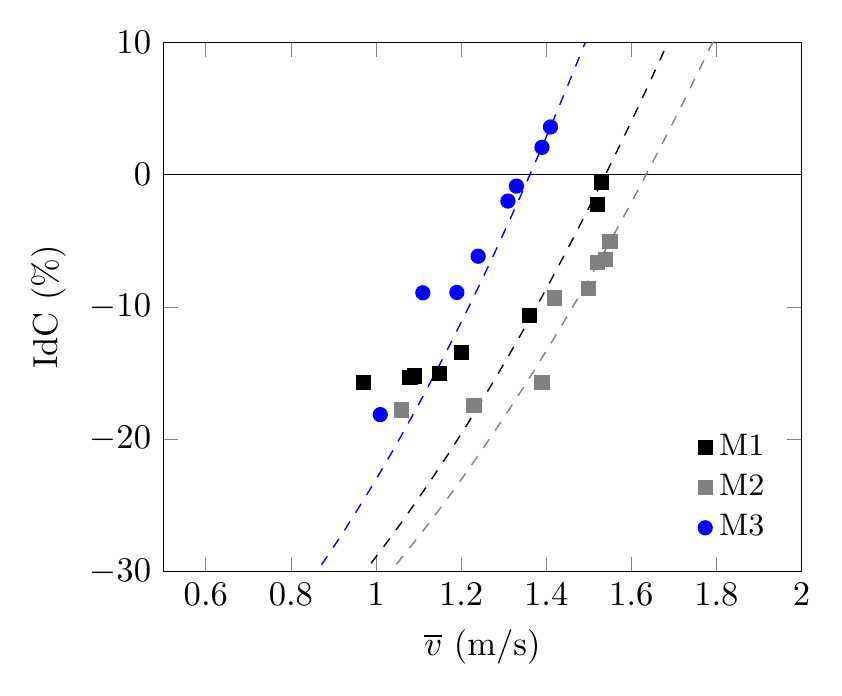}
\caption{Arm coordination with the mean velocity for 3 swimmers. The dashed lines correspond to the eq.\ref{vmean1} with $\widetilde{v} = v^*$.
}\label{IdCVDim}
\end{figure}

\begin{figure}[h!]
\centering
\includegraphics[width=.8\linewidth]{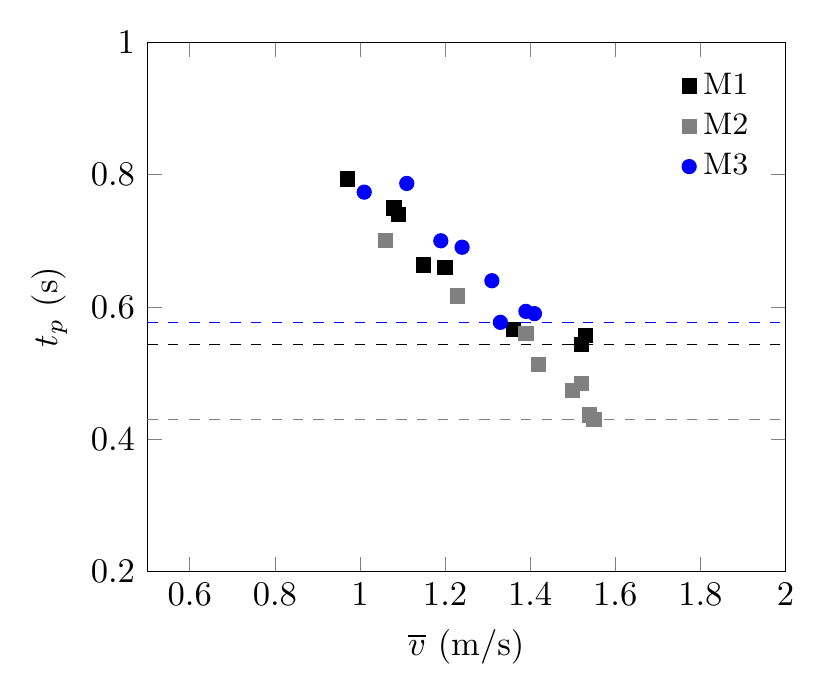}
\caption{Propulsion time with the mean velocity for the 3 swimmers presented in figure \ref{IdCVDim}. The vertical dashed lines show the minimum propulsion time achieved by the swimmers. 
}\label{taupVDim}
\end{figure}

Figure \ref{IdCVDim} shows examples of the evolution of the index of coordination with the mean velocity $\overline{v}$ for three swimmers.  Overall, it is observed that the swimmers tend to increase their coordination index as they increase their velocity. The maximum mean velocities of the swimmers M1 and M2 are close to 1.5 \si[per-mode=symbol]{\metre\per\second} but yet their coordination patterns are different, respectively -0.5\% and -5\%. On the other hand, it can be seen that the swimmers M3 and M1 are close to the opposition mode ($\IdC=0$) for drastically different velocities. At lower velocities, these three swimmers choose similar coordination patterns. This outlines the difficulty to compare swimmers technique and to provide good advice for training and performance in competition. Figure \ref{taupVDim} shows the evolution of the propulsion time for these three swimmers with the velocity. Their propulsion time decreases as the velocity increases. Their minimum propulsion time ranges from 0.6 to 0.4 \si[per-mode=symbol]{\second}. For M1 and M3, the propulsion time seems to plateau to a lower bound as the velocity is increased as outlined by the vertical dashed lines. Note that knowing the $\IdC$ and the propulsion time $t_p$, we can get the stroke rate.
 
A second test consists of graded speed test on the so-called MAD-system \cite{toussaint1988active} and enables to Measure the Active Drag of the swimmers. In this test, the swimmers push off from fixed pads spaced 1.35 \si[per-mode=symbol]{\metre} and 0.8 \si[per-mode=symbol]{\metre} below the water surface with each stroke. The system enables to estimate the drag force assuming constant mean swimming velocity \cite{truijens2005biomechanical,toussaint1988active,seifert2010}. All the swimmers were tested on 10 different speeds on the MAD-system.

\begin{figure}[h!]
\centering
\includegraphics[width=.8\linewidth]{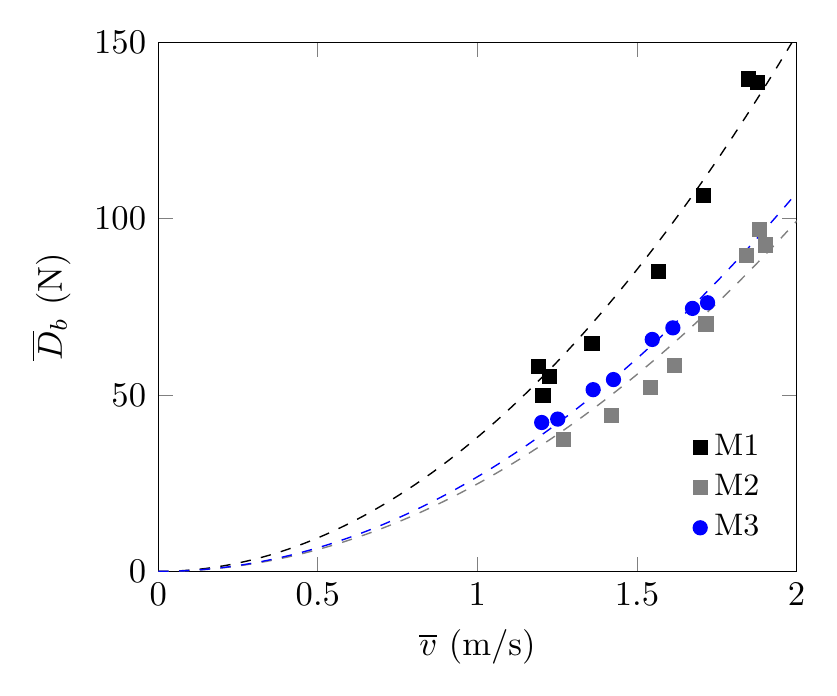}
\caption{Body drag estimated with the MAD-system with the mean velocity for three swimmers of the data set. The dashed lines correspond to the fitted curve $\overline{D}_b = k_b^\text{MAD} \overline{v}^2$.
}\label{DragVDim}
\end{figure}

Figure \ref{DragVDim} shows the obtain results for the 3 selected swimmers. The drag force is fitted to $\overline{D}_b = k_b^\text{MAD} \overline{v}^2$ to estimate the body drag coefficient. For the current three swimmers, the value range from $k_b^\text{MAD}$ = 24.8 to  38.0 \si[per-mode=symbol]{\kilogram\per\metre}. The mean value on the 16 swimmers was 30 kg/m.

To sum up, for all the 16 swimmers, we have collected information on their swimming technique (arm coordination, propulsion time) and also body characteristics (body mass, height, body drag coefficient during swimming). The goal is now to identify non-dimensional numbers enabling a fair comparison of these expert swimmers and provide a physical discussion to predict the optimal coordination. To this end, we first propose a simple model of a swimmer. 

\subsection*{Maximum force model}
Writing Newton's second law on the swimmer system in the direction of the race, we get :
\begin{align}
 \left(m_0 + m_a\right) \frac{\mathrm{d} v}{\mathrm{d}t} = T_{b} - D_{b},
\end{align}
where $m_0$ is the mass of the swimmer, $m_a$ denotes the added masses due to the acceleration of the water\footnote{In all the applications $m_a=0$ as we did not evaluate it.}, $v$ is the instantaneous velocity, $T_{b}$ the total instantaneous thrust generated by the swimmer and $D_{b}$ the body drag. Averaging on a stroke cycle and assuming a periodic regime is reached, it comes: 
\begin{align}
 0 = \overline{T}_{b}- k_b \overline{v}^2,
 \label{eqmean1}
\end{align}
where we assumed $\overline{D}_b = k_b \overline{v}^2$ and the overline denotes the average on a cycle. For instance for a quantity $a$: $\overline{a} = 1/T \int_0^T a\left(t\right)\mathrm{d}t$. 
We can further separate the thrust of each arm and define $T_a^{i}\left(t\right)$ as the instantaneous thrust of the arm $i$ at $t$. In this simplified symmetrical model, on a cycle, the two arms will produce the same mean thrust and thus we can define:
\begin{align}
\widetilde{T} _a=  \frac{1}{t_p} \int_0^{t_p} T_a\left(t\right)\mathrm{d}t,
\end{align}
where $\widetilde{T}_a$ denotes the mean thrust generated by one arm during the propulsion. It is reasonable to assume that this thrust can be controlled by the swimmer and is bounded, $\widetilde{T}_a\in[0,T_a^*]$. $T_a^*$ corresponds to the maximum thrust they can generate. Injecting this in eq.\ref{eqmean1}, we get:
\begin{align}
 0 = \frac{2t_p}{T}\widetilde{T}_a - k_b \overline{v}^2.
 \label{eqmean2}
\end{align}

\begin{figure}[h!]
\centering
\includegraphics[width=.8\linewidth]{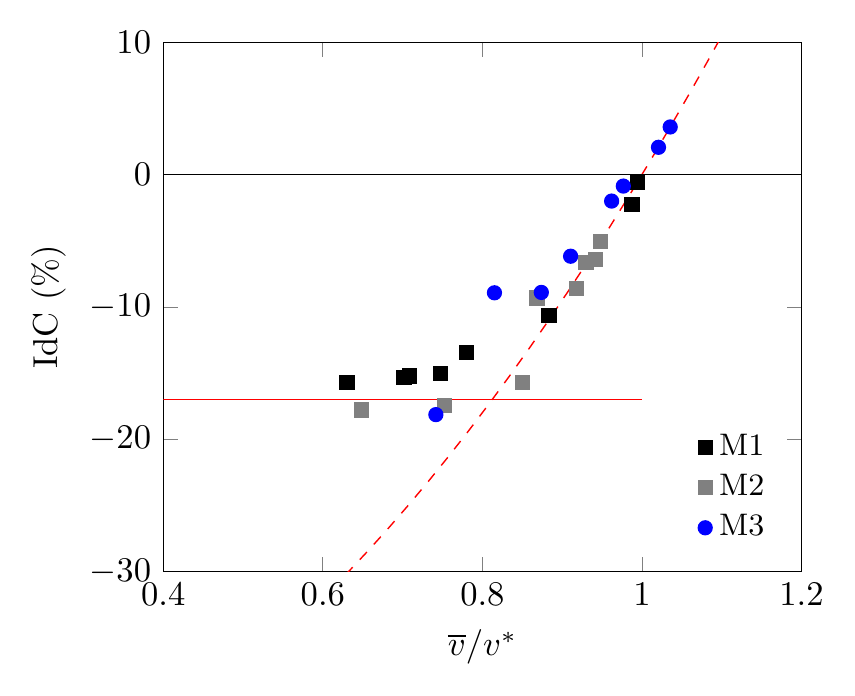}
\caption{Arm coordination with the mean velocity for 3 swimmers of the data set. The dashed lines correspond to the eq.\ref{vmean1}. The solid red line is a guide for the eyes outlining the plateau the swimmers seem to converge toward.
}\label{IdCVNoDim3swimmers}
\end{figure}
 
Using the fact that $T=2t_p-2t_c$, we get $2t_p/T = 1+2\IdC$. We then find a relationship between the coordination index and the velocity, which depends on the mean thrust generated by one arm during its propulsive phase and the body drag: 
\begin{align}
\overline{v} = \widetilde{v}\left(1+2 \IdC\right)^{\sfrac{1}{2}},
\label{vmean1}
\end{align}
where $\widetilde{v} = \sqrt{{\widetilde{T}_a}/{k_b}}$. This basic discussion enables to define a characteristic velocity $v^*$ which depends on the swimmers mean maximum thrust, $T_a^*$ and their body drag coefficient $k_b$. It also appears in this simple model that to swim faster the swimmers can play on their coordination once they maximize their thrust, assuming constant body drag coefficient $k_b$. We can use this model to characterize the swimmer velocity. It can be assumed that at the maximum velocity of the previous coordination test (see figure \ref{IdCVDim}) the expert swimmers used their maximum thrust to produce their highest speed. It comes:
\begin{align}
v^* = \frac{\overline{v}_{\max} }{\sqrt{1+2 \IdC_{\max}}},
\label{vstart}
\end{align}
where the index $\max$ denotes the test with maximum velocity for the swimmer. The dashed lines in figure \ref{IdCVDim} correspond to the eq.\ref{vmean1} with $\widetilde{v} = v^*$ and is referred to as the maximum force model. The swimmers follow nicely the model of maximum force when their velocity increases. It is observed that as they simulate longer races (lower velocities) they tend to diverge from this simple maximum force model and seem to use less thrust. This is in agreement with the observations of Craig \& Pendergast \cite{craig1979relationships}. This method of characterization of the velocity is applied to the three swimmers presented previously. The results are displayed in figure \ref{IdCVNoDim3swimmers}. It appears that the swimmers use similar coordination at similar non-dimensional velocities. We observe two swimming strategies: one following the red dashed line and corresponds to the maximum force model and one with a rather constant index of coordination (solid red line).

\subsection*{Non-dimensional velocity and coordination}

\begin{figure}[h!]
\centering
\includegraphics[width=.8\linewidth]{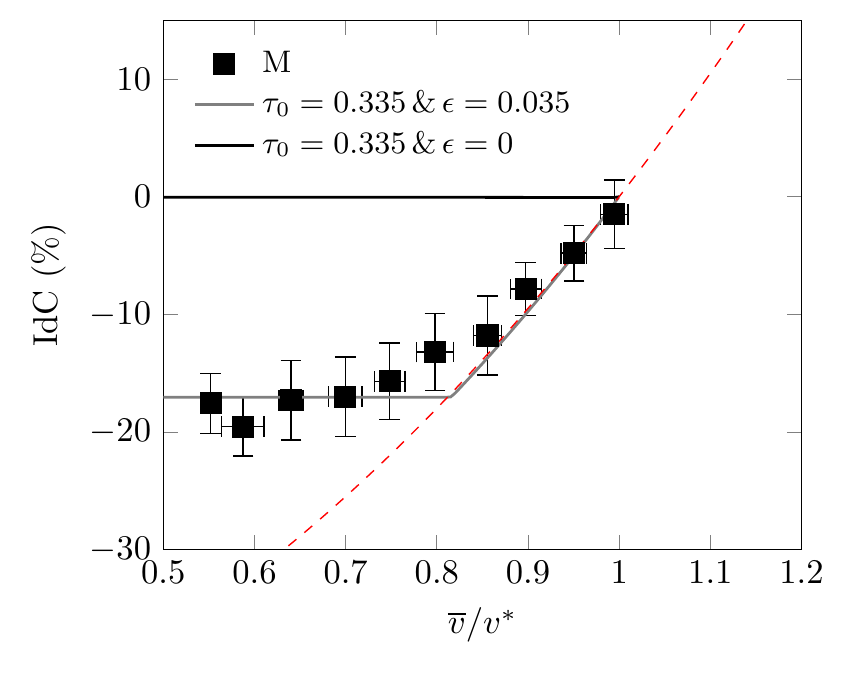}
\caption{Arm coordination with the mean non-dimensional velocity for the 16 swimmers of the data set. The dashed line corresponds to the maximum force model. The solid lines show the optimal coordination with $\tau_0=0.335$ and $\epsilon=0$ and $\epsilon=0.035$ in black and gray, respectively.
}\label{IdCVNoDim}
\end{figure}

We apply this analysis on the 16 expert swimmers tested in 2007. In all the cases, we use their maximum velocity to define their characteristic velocity $v^*$ (see eq.\ref{vstart}). We group the swimmers in pools of similar $\overline{v}/v^*$ with steps of 0.05 and averaged the data. Each pool contains at least 4 points and 4 different swimmers. In average, there are 14 observations per pool with 10 different swimmers. The results are displayed in figure \ref{IdCVNoDim}. This figure is one of the main results of the present paper. The data are also provided in SI appendix 1 for the separated swimmers. 

In this data set, we  observe that these expert swimmers follow nicely the maximum force model presented in the previous section for non-dimensional velocity higher than 0.8. Below this value, the index of coordination is almost constant and near a value of -15\% --  -20\%. 

For sprint races, the expert swimmers do not have to worry about their energy consumption and should maximize their velocity. This is achieved by using the maximum thrust and the highest reachable index of coordination. On the contrary, for mid and long distance races, the expert swimmers need to manage their energy consumption and adopt the index of coordination that enables them to maximize the distance of the race they can swim maintaining this velocity.
The race distance they can reach depends on physiological data such as their maximum rate of energy supplied by the oxygen and their anaerobic reserve \cite{keller1973ia,keller1974optimal,behncke1987optimization}.
In other words, an expert swimmer should adopt the arm coordination that minimizes the energy cost at a given velocity for mid and long distance races \cite{ di1974energetics,BARBOSA2010, leblanc2007intra}. 
In the next section, we propose a simple model to understand the observed coordination at low velocities (hence simulating long races) and predict the index of coordination plateau for a given swimmer depending on their physical characteristics.

\section*{Burst-and-coast in catch-up mode}
\subsection*{Physical model}
Burst-and-coast swimming behavior is quite common in nature. It consists of cyclic burst of swimming movements followed by gliding phase in which the body is not producing thrust. This surprising strategy of propulsion is observed in fishes such as cod and saithe and was shown to be actually energetically cheaper than steady swimming at the same average velocity \cite{videler1982energetic}. Mathematical model helped understood this non-intuitive behavior \cite{weihs1974energetic} where non-continuous propulsion could be cheaper. In the present paper, the model proposed by Weihs \cite{weihs1974energetic} and Videler and Weihs \cite{videler1982energetic} is adapted to human swimming with the arms only.

The main assumption in the model is that the resistance is not the same during active and passive swimming. 
In the catch-up mode of coordination, the swimmers alternate between phases with active propulsion and phases of gliding. During active swimming the water resistance will be assumed to have the form $D_b = k_b v^2$ and the swimmer produces a thrust $T_b$, which will be considered constant (in order to keep the model simple).  For the gliding phase, the swimmers have one arm forward fully extended (similarly to the figure \ref{diffLongSprint}-a). The drag should be reduced and will be modeled by $D_b = (1-\epsilon) k_b v^2$, where $\epsilon\geq0$. Note that if $\epsilon<0$ then clearly, the swimmers should never try to glide as their resistance is greater during this phase (this could be the case for non expert swimmers). The parameter $\epsilon$ denotes the gliding effectiveness of the swimmer and is part of swimming technique. As the stroke is supposed periodic and the two arms symmetrical, we limit the study to half a stroke cycle. In other words, we focus on a single arm. The swimmer velocity oscillates between two extreme values denoted $\vmin$ and $\vmax$. The equations to solve can be written as:
\begin{align}
 \left(m_0 + m_a\right) \frac{\mathrm{d} v}{\mathrm{d}t} &= T_{a} - k_b v^2, & 0&\leq t\leq t_p,
  \label{eqDim1}
 \\
 \left(m_0 + m_a\right) \frac{\mathrm{d} v}{\mathrm{d}t} &=  - (1-\epsilon) k_b v^2,  & t_p&\leq t\leq T/2,
 \label{eqDim2}
 \end{align}
 with the boundary conditions: 
 \begin{align}
 v(0) &= v(T/2) = \vmin,\\
 v(t_p) & =\vmax.
\end{align}
Similarly to the previous section, we will consider that the swimmers can control their thrust and that it is bounded $T_a\in\left[0, T_a^*\right]$. An extension to superposition is discussed in SI appendix 2.

\begin{figure}[h!]
\centering
\includegraphics[width=0.9\linewidth]{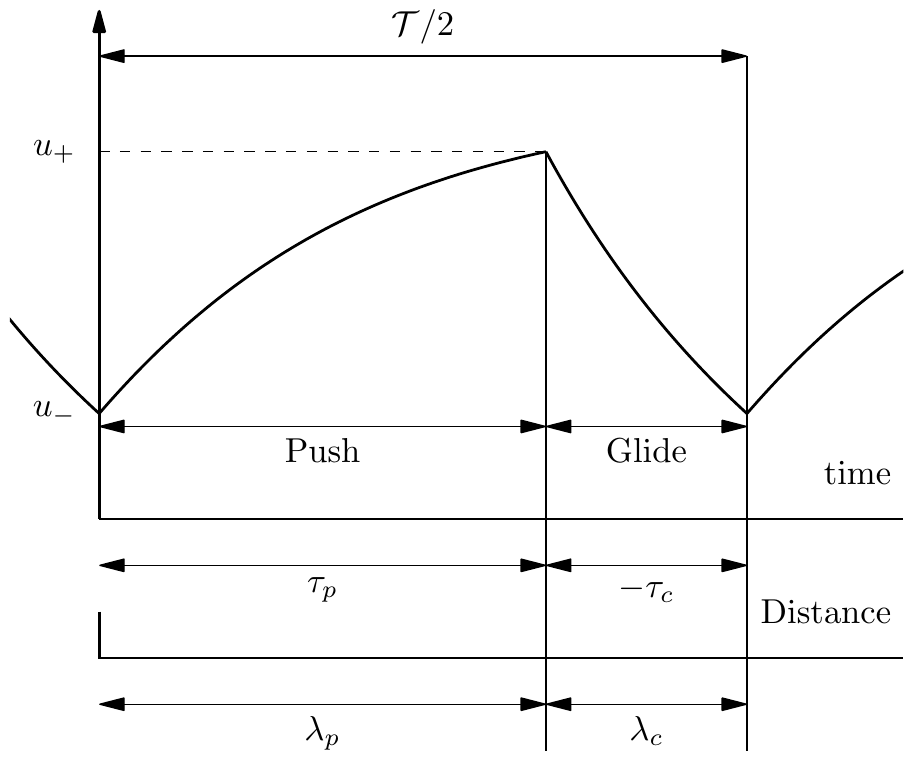}
\caption{Burst-and-coast model for swimmers intra-cycle velocity variations and notations.
}\label{SketchPushAndGlide}
\end{figure}

To non-dimensionalize this set of equations, we define $\tau = t/\tau^*$, $u = v/v^*$ and $\varphi = T_a/T_a^*$, where $\tau^*$, $v^*$ are the characteristic time and velocity defined by:
\begin{align}
\tau^* = \frac{m_0+m_a}{k_b v^*},\label{tauseq}\\
v^* = \sqrt{T_a^*/k_b}.\label{vseq}
\end{align} 
Note that $v^*$ is the same as the one defined in the previous simple model. Using these definitions, we rewrite the eq.\ref{eqDim1}--\ref{eqDim2} in the form:
\begin{align}
\frac{\mathrm{d} u}{\mathrm{d}\tau} &= \varphi - u^2, & 0&\leq \tau\leq \tau_p,
  \label{eqNoDim1}
 \\
  \frac{\mathrm{d} u}{\mathrm{d}\tau} &=  - (1-\epsilon) u^2,  & \tau_p&\leq \tau \leq \mathcal{T} /2,
 \label{eqNoDim2}
 \end{align}
 with boundary conditions:
  \begin{align}
 u(0) &= u(\mathcal{T}/2) = \umin,\\
 u(\tau_p) & =\umax.
\end{align}
Figure \ref{SketchPushAndGlide} shows half a cycle with the different notations.

One of the main motivations behind this simple model is that it can be solved analytically. It is rather simple to show that:
\begin{align}
\umax &= \sqrt{\varphi} \tanh \left[ \tau_p\sqrt{\varphi} + \tanh^{-1} \left(\frac{\umin}{\sqrt{\varphi}}\right) \right],\label{umaxeq}\\
\tau_c &= \frac{1-\umax/\umin}{(1-\epsilon) \umax},\\
\lambda_p &= \log\left(\sqrt{\dfrac{1-\umin^2/\varphi}{1-\umax^2/\varphi}}\right),\\
\lambda_c &= \dfrac{1}{1-\epsilon}\log\left(\dfrac{\umax}{\umin}\right),\label{lambdaceq}
\end{align}
where $\lambda_p$ ($\lambda_c$) is the non-dimensional distance travelled during the propulsive (gliding) phase (respectively). It is important to outline that $\tau_c$ is the non-dimensional coordination time and is negative in catch-up mode. The mean velocity can be evaluated as:
\begin{align}
\overline{u} = \dfrac{\lambda_p+\lambda_c}{\tau_p-\tau_c}.
\end{align}

Then the objective of expert swimmers is to swim a given distance in the minimum time. To achieve this goal for long distance races, they have to manage their energy consumption. At a given mean velocity, they should select the coordination that enables them to minimize their propulsion cost.
This will enable them to swim the longest distance at this mean velocity.
The energy consumed during catch-up mode is the energy expended solely during the propulsion phase:
\begin{align}
\mathcal{E}_\text{cu} &= \int_0^{\tau{_p}} \varphi u \mathrm{d}\tau, \nonumber \\
			  & = \varphi \lambda_p.
\end{align}
For this amount of energy, the swimmers travel a non-dimensional distance $\lambda_p+\lambda_c$. To travel the same distance in opposition mode (steady swimming), they would use: 
\begin{align}
\mathcal{E}_\text{op,cu} &= \int_0^{\mathcal{T}/2} \overline{\varphi}\, \overline{u} \mathrm{d}\tau, \nonumber \\
			  & = \overline{\varphi} \left(\lambda_p+\lambda_c\right),
\end{align}
where $\overline{\varphi}$ is the thrust required to swim at the constant velocity $\overline{u}$ and using eq.\ref{eqNoDim1} it comes:
\begin{align}
\overline{\varphi} = \overline{u}^2. 
\end{align}
The economy can therefore be defined as: 
\begin{align}
\mathcal{R}_\text{cu}=\dfrac{\mathcal{E}_\text{cu} }{\mathcal{E}_\text{op,cu}} =   \dfrac{\varphi}{\overline{u}^2} \dfrac{\lambda_p}{\lambda_p+\lambda_c}.
\end{align}
Most of the development are similar to the one in the paper of Videler and Weihs \cite{videler1982energetic}. We further add the assumption that the propulsion phase duration $\tau_p$ is limited by the arms dynamic. Note that in superposition mode, the economy is always larger than 1. Therefore, for $u<1$ the swimmer should at worst prefer the opposition mode. Above $u>1$, superposition is the only possible mode of coordination. An extension of the present model to the superposition mode of coordination is discussed in SI appendix 2.

\subsection*{Propulsion time assumption}
In human swimming, the propulsion phase duration corresponds to the time the hand needs to travel from the fully extended forward position to the release from the water near the hips. To estimate this time of propulsion, we will simplify the arm+hand to a simple paddle, which travels twice the arm length on a straight line at constant velocity (we neglect the acceleration phases). The thrust generated by the swimmer solely comes from this paddle. The propulsion time is then:
\begin{align}
\tau_p \approx \dfrac{2 \lambda_a}{u_{h/b}},
\end{align}
where $\lambda_a$ is the non-dimensional arm length and $u_{h/b}$, the non-dimensional hand velocity in the body frame. As the hand travels much faster than the body through the water, we neglect the contribution of the body velocity. We then expect the propulsion time to be of the order of:
\begin{align}
\tau_p \approx  \dfrac{2 \lambda_a}{u_{h/w}},
\end{align}
where $u_{h/w}$ is the hand velocity with respect to the water. This velocity depends on the resistance coefficient of the hand $\alpha_h = k_h/k_b$ and the force used by the swimmer to move it through the water\footnote{we discuss the impact of the arm speed on the achievable force in SI appendix 3}:
\begin{align}
\alpha_h u_{h/w}^2 = \varphi,
\label{handseq}
\end{align}
and therefore:
\begin{align}
\tau_p \approx  \dfrac{\tau_0}{\sqrt{\varphi}},
\end{align}
where $\tau_0$ is a a characteristic time of propulsion. It depends on the arm length $\lambda_a$ and the $\alpha_h$ coefficient:
\begin{align}
\tau_0 \approx 2\lambda_a \sqrt{\alpha_h}.
\label{tau0exp}
\end{align}
The propulsion time increases with the arm length (larger distance to travel) and the hand size ($\alpha_h$ increases). Note that it does not depend on the swimmer's force. Using the dimensional parameters, we get $\tau_0 \approx 2 L_a \sqrt{k_h k_b}/(m_0+m_a)$, with $L_a= 0.6$ \si[per-mode=symbol]{\metre} the arm length, $m_0=80$ \si[per-mode=symbol]{\kilogram}, $k_h\approx 13$ \si[per-mode=symbol]{\kilogram\per\metre} the drag coefficient of the hand \cite{martin1981simple} and $k_b = 30$ \si[per-mode=symbol]{\kilogram\per\metre} . This gives $\tau_0 \approx 0.30$.

\begin{figure}[h!]
\centering
\includegraphics[width=.8\linewidth]{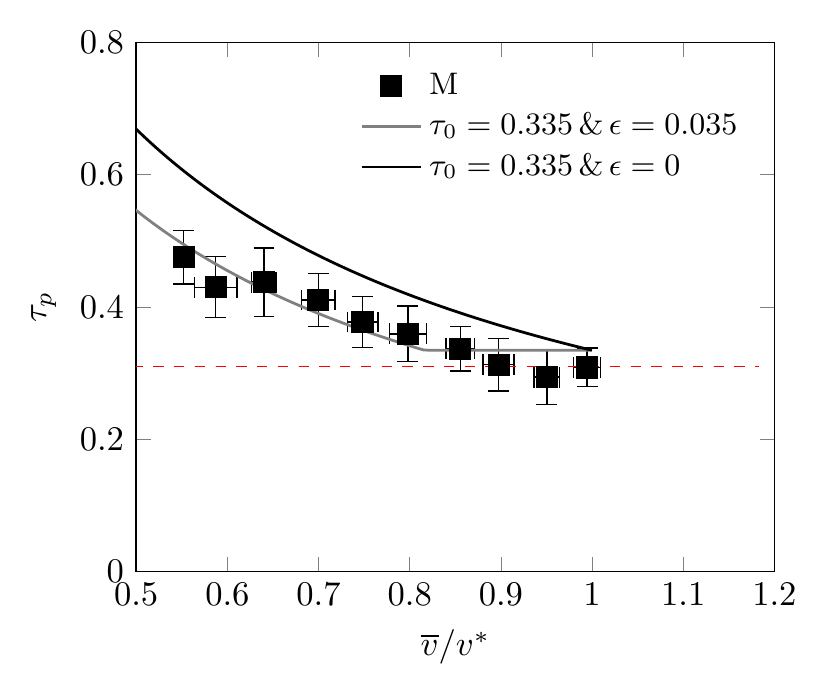}
\caption{Non-dimensional Propulsion time of expert swimmers with the non-dimensional mean velocity. The dashed red lines show the evaluation of the $\tau_0$ based on the observation. The solid lines show the optimal propulsion time with $\tau_0=0.335$ and $\epsilon=0$ and $\epsilon=0.035$ in black and gray, respectively. 
}\label{TaupVNoDim}
\end{figure}
 
To verify this assumption, we look at the propulsion time of our swimmers. We can evaluate the propulsion time as:
\begin{align}
\tau_p = \dfrac{t_p}{\tau^*}.
\end{align}
Figure \ref{TaupVNoDim} shows the obtained mean value of the non-dimensional propulsion phases with the non-dimensional velocity. We see that the propulsion time plateaus to $\tau_0 =0.31\pm 0.03$ when the points are on the maximum force model in figure \ref{IdCVNoDim}:
\begin{align}
\overline{u} = \dfrac{\overline{v}}{v^*} = \left(1+2\IdC\right)^{\sfrac{1}{2}}.
\end{align}
This corresponds to $\varphi=1$ and $\epsilon=0$. It is in good agreement with the previous estimations.

\subsection*{Comparison to the data set}
To summaries, the swimmers are characterized by two measurable parameters: the propulsion time parameter $\tau_0$ and the gliding effectiveness $\epsilon$. Both parameters could be measured on swimmers. 
To minimize their propulsion cost, they can play on their force used to produce the thrust $\varphi \in \left[0,1\right]$. We can write this problem in the form:
\begin{align}
\min_{\varphi,\, \text{s.t.}\, \overline{u} = u_0} \mathcal{R}_\text{cu}\left(\varphi\right).
\label{propR}
\end{align}
This problem can be solved and would yield the optimal coordination strategy for our model swimmer for a given mean velocity $\overline{u} = u_0$.

To compare the present burst-and-coast model to our data set, we used a Powell's conjugate direction method \cite{powell1964efficient} with Golden-section search \cite{kiefer1953sequential} on $\epsilon$ and $\tau_0$ to minimize the error on the model prediction on the index of coordination and propulsion time. The obtained best-fit parameters are $\epsilon = 0.035$ and $\tau_0 = 0.335$. Note that $\tau_0$ is in the error bar of the previous estimation. Figures \ref{IdCVNoDim} and \ref{TaupVNoDim} show the obtained best fit compared to the data set. We further added the optimal choice for the limit case $\epsilon=0$. For this case, the model predicts the intuitive choice of the opposition mode ($\IdC=0$) as the optimal choice of coordination. Indeed, if there is no benefit to glide ($\epsilon\leq0$), then the swimmer should not glide.

The model shows that there exists a single optimal coordination (here $\IdC_c \approx -17\%$) for non-dimensional velocity lower than $u_c = 0.81$. The swimmers would save $\approx1\%$ energy. It is indeed possible to be more efficient in catch-up mode at certain speed with this model, even-though this can be counter-intuitive \cite{havriluk2012expert}. Above this critical velocity, the swimmers use their maximum force and hence follow the maximum force model presented before. 

The present model also provides information on the intra-cycle velocity variation (IVV) of the swimmers. With the best-fit parameters, our swimmers have a relative velocity variation of 13\% during the $\IdC$ plateau and it decreases once the swimmers reach the maximum force model. In this simplified model, the relative velocity variation goes to zero as the $\IdC$ goes to 0 due to constant force approximation during the propulsion phase. Yet the reduction of the relative velocity variation with the mean velocity is consistent with the observations of Matsuda \textit{et al.} \cite{matsuda2014intracyclic}.

\subsection*{Linear approximation and two regimes} 
It is interesting to note that in eq.\ref{umaxeq}--\ref{lambdaceq}, we can define $X  = \umin/\sqrt{\varphi}$ and write all the parameters as a function of $X$ only. Therefore, $\mathcal{R}$ is a function of $X$ only.

\begin{figure}[h!]
\centering
\includegraphics[width=.8\linewidth]{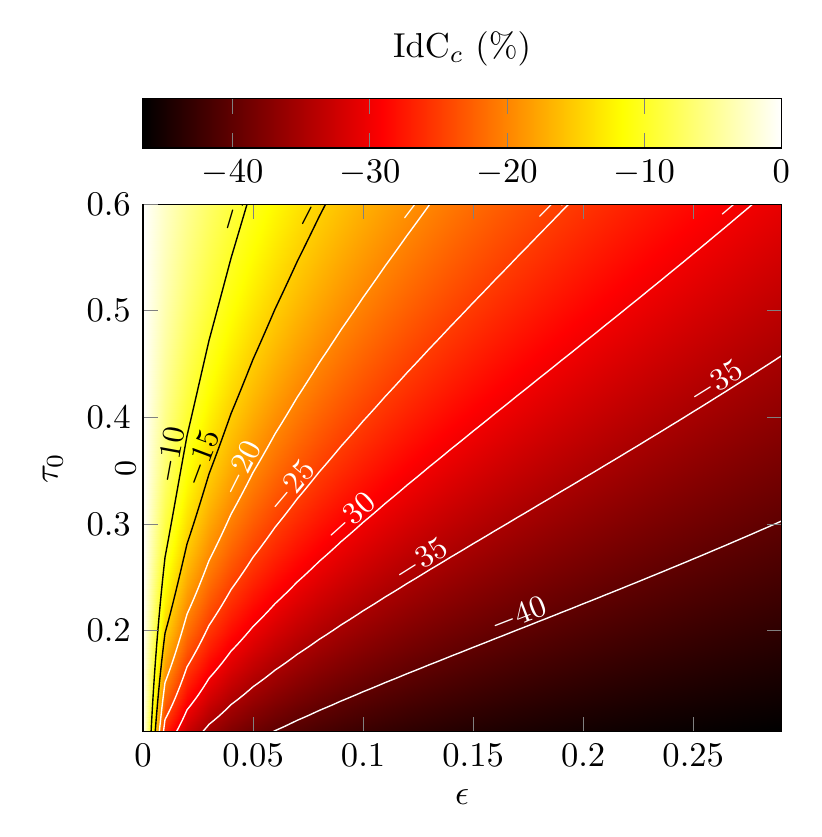}
\caption{Colormap of the optimal coordination in the Burst-And-Coast model with $\epsilon$ and $\tau_0$.
}\label{colormap}
\end{figure}

For our expert swimmers, we already observed that the maximum force model yields rather good predictions for the coordination and velocity (see red dashed line in figure \ref{IdCVNoDim}). We expect the coefficient $\epsilon$ to be close to 0 for our swimmers. We also found that $\tau_0$ was reasonably small. To linearize the equations developed in the two previous subsections, we will assume $\tau_0\ll1$ and $\epsilon \ll 1$. We further assume that $\epsilon \ll \tau_0$. Keeping only the smallest order terms in $\epsilon$ and $\tau_0$, it comes:
\begin{align}
\mathcal{R}_\text{cu}\left(X\right) \approx 1 -  \dfrac{{\tau_0}^2}{2}  \left(1- \dfrac{1+{X}^4}{2{X}^2}\right) +\epsilon \left({X}^2-1\right).
\end{align}
This approximated function has a single minimum in $X\in\left[0,1\right]$:
\begin{align}
X_0 = \dfrac{1}{\left(1+4\epsilon/\tau_0^2 \right)^{\sfrac{1}{4}}},
\end{align}
and its value is:
\begin{align}
\mathcal{R}_\text{cu}^{\min} = 1-\epsilon+ \dfrac{\tau_0^2}{2} \left(\sqrt{1+4\epsilon/\tau_0^2 }-1\right).
\label{Rmincu}
\end{align}

Injecting this expression in the evaluation of the index of coordination and keeping only the first order term, it comes that the optimal index of coordination is:
\begin{align}
\IdC_c \approx -\dfrac{1}{2} \left(1 - \dfrac{1}{\sqrt{1+4\epsilon/\tau_0^2}}\right).
\label{IdC0Exp}
\end{align}
This will be the optimal choice of coordination as long as $\overline{u} < \left(1+2\IdC_c\right)^{\sfrac{1}{2}}$. Above the critical speed:
\begin{align}
u_c \approx \dfrac{1}{\left(1+4\epsilon/\tau_0^2 \right)^{\sfrac{1}{4}}},
\label{uceq}
\end{align}
the swimmers will then switch to the maximum force regime. This is in good agreement with the observations of Craig \& Pendergast \cite{craig1979relationships}. Figure \ref{colormap} shows a colormap of the numerically found optimal index of coordination with $\tau_0$ and $\epsilon$.

From eq.\ref{IdC0Exp}, we observe that the optimal index of coordination decreases (increases) when $\epsilon$ ($\tau_0$) increases. It is not surprising to expect that the swimmer will tend to glide more when $\epsilon$ increases. Note that if $\epsilon=0$, then the expression yields that the optimal coordination is the opposition mode.  
The effect of varying $\tau_0$ is maybe less obvious. Increasing $\tau_0$ (keeping all the other parameters constant) can be compared to swimming with paddles. This will increase the parameter $\alpha_h$ in eq.\ref{tau0exp}. From the present model, we then expect the swimmers to change their coordination pattern toward the opposition mode ($\IdC$ closer to zero) with paddles. This prediction is observed by Sidney \textit{et al.}\cite{sidney2001effect}. 

In all the results, presented so far the legs were tied and could not be used by the swimmers. We discuss their effects in SI appendix 4. It is rather clear that allowing the legs to kick will lead to an increase of the swimmer characteristic velocity $v^*$. This will affect the non-dimensional propulsion time $\tau_0$. If we suppose that the legs do not affect the gliding effectiveness $\epsilon$, the coordination is then affected accordingly by an increase of the $\IdC_c$. We predict an increase of the $\IdC_c$ from -17\% to -12\% with this assumption and observe it on a group of similar level swimmers.
 
\section*{Conclusion and applications}
In the current study, we tried to understand how the arm coordination patterns of 16 expert front crawl swimmers vary according to the active drag performing two tests: an incremental coordination test, where swimmers were requested to simulate their racing techniques at 8 different speeds and a drag measurement test to measure their active drag using the MAD-system. In both tests, the swimmer legs where tied and they were equipped with a pull-buoy to avoid that their legs sank. To compare the evolution of the different swimmer coordination ($\IdC$) with their mean velocity ($\overline{v}$), we defined a characteristic velocity $v^* = \overline{v}_{\max}/\sqrt{1+2\IdC_{\max}}$ using their maximum velocity test. We observed that swimmers used similar coordination patterns for a given non-dimensional velocity $\overline{u} = \overline{v}/v^*$. At low $\overline{u}$, the swimmers seem to select a constant negative index of coordination and above a critical non-dimensional velocity of about 0.8 their coordination increases with their velocity. To further understand these two regimes, we propose a physical model of burst-and-coast \cite{weihs1974energetic} adapted to swimming in front-crawl. The main idea of this model is that the swimmers experienced a reduced drag while gliding with one arm extended forward. We compare this drag to the active swimming drag during the underwater stroke by defining a gliding effectiveness coefficient $\epsilon>0$. One additional key assumption in our model is the arm propulsion time. We proposed that this time depends on the force used to propel the body through the water and is bounded by a lower value $\tau_0$ which depends on the swimmer characteristics only. We then showed that this model predicts that there exist two swimming regimes similarly to the observations.

A "low velocity" regime, where the swimmers select a constant index of coordination and reduce their propulsion force to minimize their propulsion cost. A "high velocity" regime, where the swimmers increase their index of coordination to push at maximum force and to gain more speed. It is in this latter regime that the transition from catch-up to superposition can occur.

The optimal index of coordination in the low velocity regime is $\IdC_c \approx -(1-1/\sqrt{1+4\epsilon/\tau_0^2})/2$ and the transition from the "low velocity" to "high velocity" regime will occur at $u_c \approx 1/\left(1+4\epsilon/\tau_0^2\right)^{\sfrac{1}{4}}$ in the limit $\tau_0\ll 1$ and $\epsilon\ll \tau_0$. This transition can be linked to swimming distances through an energetic equation. This transition is similar to the one observed by Craig \& Pendergast \cite{craig1979relationships} in between the 200 m and 400 m races (see SI appendix 5).

Using this model, it is possible to advice on individual optimal arm coordination  for each swimmer based on his/her physical characteristics. To predict their optimal coordination, we need to evaluate the two parameters $\epsilon$ and $\tau_0$. It is possible to use a gliding test to estimate the value of $\epsilon$ by varying the arm position of the swimmer at the surface. For the propulsion time parameter, one could evaluate the time to perform a single arm pull with the maximum possible thrust on a 25m sprint with index of coordination measurements. Then $\tau_0 = t_0 k_b v^*/m$, where $k_b$ would be the drag measured from the previous gliding test with the arms along the body. These tests could be also done on swimmers with disabilities and advice them on individual optimal arm coordination based on their physical characteristics and type of impairment.

\section*{Acknowledgment}
The authors would like to thank all the athletes that participated in the tests. They are also grateful to Leo Chabert, Benoit Bideau and Vincent Bacot for their help at different stages of the project and the useful discussions. Last but not least, the authors would like to thank The Olympic Multimedia Library for granting us access to the footage of the race and allowing us to use the images to illustrate our work.

\bibliographystyle{abbrv}
\bibliography{bibfile}

\section*{Supplementary Information}
\subsection{Index of coordination of the 16 swimmers}
\label{separatedSwimmer}
\begin{figure*}[h!]
\centering
\includegraphics[width=.5\linewidth]{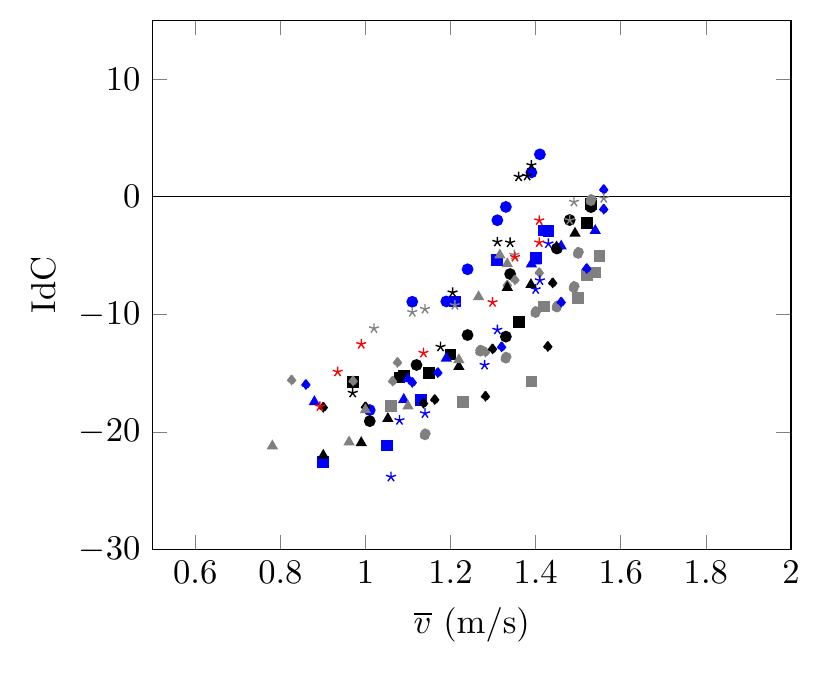}\includegraphics[width=.5\linewidth]{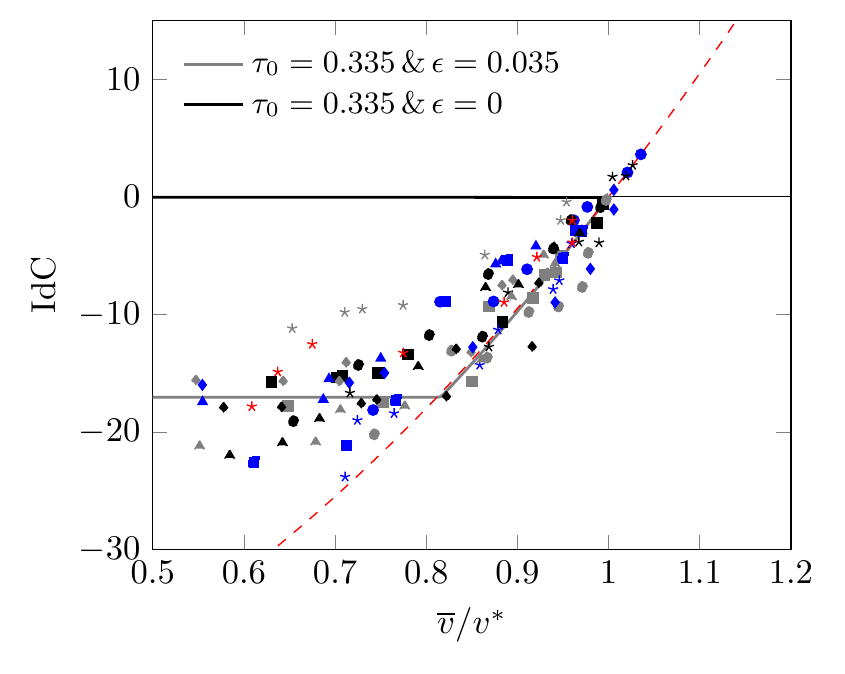}
\caption{Arm coordination with the velocity for the 16 swimmers separately. Each symbol represents a single swimmer. The dashed correspond to the maximum force model while the solid lines show the result of the model presented in the paper for the parameters listed. 
}\label{IdCVDIMALL}
\end{figure*}

We provide here the raw data for all the swimmers for the index of coordination with dimensional velocity and after applying the evaluation of the characteristic speed $v^* =\overline{v}_{\max}/\sqrt{1+2\IdC_{max}}$, where $\overline{v}_{\max}$ denotes the maximum velocity achieved by the swimmer. The results are displayed in figure \ref{IdCVDIMALL}.

\subsection{Extension to superposition mode}
\label{extensionSuperposition}
The burst-and-coast model can be extended to the superposition mode. In this case, we have the following equations:
\begin{align}
\frac{\mathrm{d} u}{\mathrm{d}\tau} &= 2\varphi - u^2, & 0&\leq \tau\leq \tau_c,
  \label{eqNoDim1Ext}
 \\
  \frac{\mathrm{d} u}{\mathrm{d}\tau} &=  \varphi-  u^2,  & \tau_c&\leq \tau \leq \mathcal{T} /2,
 \label{eqNoDim2Ext}
 \end{align}
 with boundary conditions:
  \begin{align}
 u(0) &= u(\mathcal{T}/2) = \umin,\\
 u(\tau_c) & =\umax,
\end{align}
where $u = v/v^*$ is the non-dimensional velocity, $\tau = t/\tau^*$ the non-dimensional time, $\varphi = T_a/T_a^*$ the non-dimensional thrust and:
\begin{align}
\tau^* = \frac{m_0+m_a}{k_b v^*},\label{tauseq}\\
v^* = \sqrt{T_a^*/k_b}.\label{vseq}
\end{align}
Note that $v^*$ is the same as the one defined in the previous appendix.
 
\begin{figure}[h!]
\centering
\includegraphics[width=.8\linewidth]{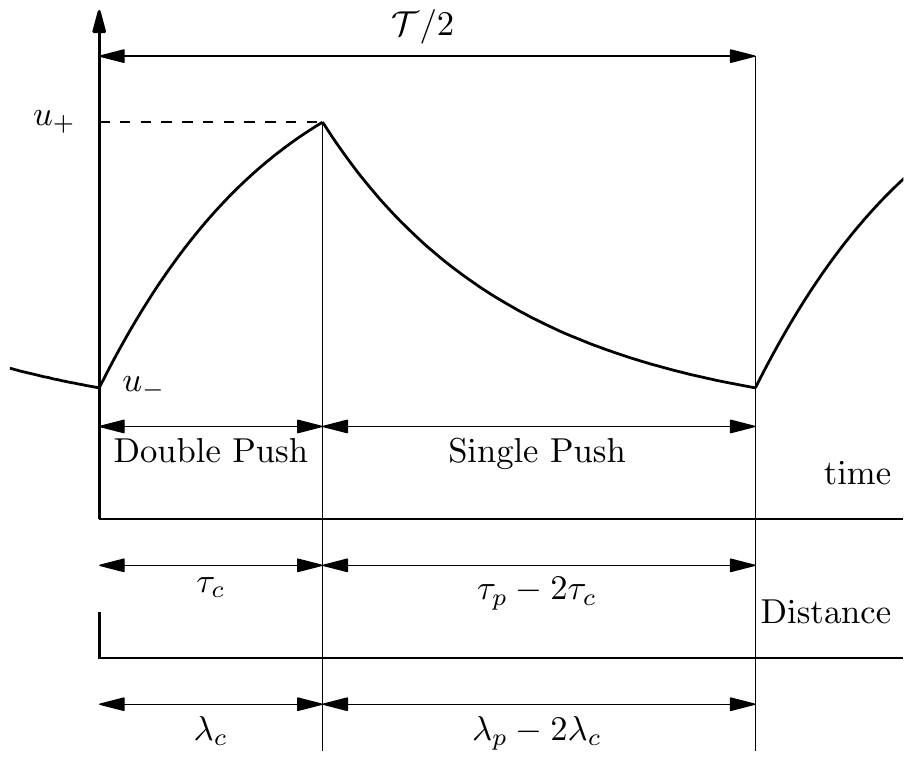}
\caption{Burst-and-coast model for swimmers intra-cycle velocity variations and notations extension to the superposition mode. 
}\label{SketchPushAndGlideSuper}
\end{figure}
 
Here, the propulsion time does not appear directly in the equations. We still have the relation:
\begin{align}
\mathcal{T}/2 = \tau_p-\tau_c.
\end{align}
Figure \ref{SketchPushAndGlideSuper} summarizes the notations. This system can also be solved analytically:
\begin{align}
\umax &= \sqrt{2\varphi} \tanh\left[\tau_c \sqrt{2\varphi} + \tanh^{-1} \left(\dfrac{\umin}{\sqrt{2\varphi}}\right) \right],\\
\tau_p - 2\tau_c &= \dfrac{1}{\sqrt{\varphi}} \left\{\coth^{-1}\left(\dfrac{\umin}{\sqrt{\varphi}}\right)-\coth^{-1}\left(\dfrac{\umax}{\sqrt{\varphi}}\right)  \right\},\\
\lambda_c &= \log\left(\sqrt{\dfrac{1-\umin^2/2\varphi}{1-\umax^2/2\varphi}} \right),\\
\lambda_p-2\lambda_c &= \log\left(\sqrt{\dfrac{\umax^2/\varphi-1}{\umin^2/\varphi-1}} \right).
\end{align}
The energy consumed during the superposition mode is:
\begin{align}
\mathcal{E}_\text{su} &= \int_0^{\tau_c} 2\varphi u\mathrm{d}\tau + \int_{\tau_c}^{\tau_p} \varphi u\mathrm{d}\tau
\\
			  &= \varphi  \lambda_p.
\end{align}
For this amount of energy, the swimmers travel a non-dimensional distance $\lambda_p-\lambda_c$ only. It should be compared to the energy used to travel the same distance in opposition mode (steady swimming):
\begin{align}
\mathcal{E}_\text{op,su} &= \int_0^{\mathcal{T}/2} \overline{\varphi}\, \overline{u} \mathrm{d}\tau\\
			  &= \overline{u}^2 \left(\lambda_p-\lambda_c\right).
\end{align}
The economy can therefore be defined as:
\begin{align}
\mathcal{R}_\text{su} =  \dfrac{\mathcal{E}_\text{su}}{\mathcal{E}_\text{op,su}} = \dfrac{\varphi}{\overline{u}^2} \dfrac{\lambda_p}{\lambda_p-\lambda_c}.
\end{align}
We still assume that $\tau_p = \tau_0/\sqrt{\varphi}$ as each arms have their own control.

Defining  $X_- = \umin/\sqrt{\varphi}$ and $X_+ = \umax/\sqrt{\varphi}$, we have:\begin{align}
1<X_-<X_+<\sqrt{2}.
\end{align}
Then, assuming $X_+ = X_- +\Delta X$, it is possible to show that in the limit $\tau_0\ll 1$ and $\Delta X\ll 1$:
\begin{align}
\mathcal{R}_\text{su} \approx 1+\left[\dfrac{\left(X_- -1\right) \left(X_- +1\right) \left({X_-}^2-2\right)}{4 {X_-}^3} \right]^2\tau_0^2>1.
\end{align} 
Therefore, as expected intuitively, this mode is always more expensive energetically than the opposition. This mode should be use solely for sprinting and at maximum force ($\varphi=1$). It is interesting to note that $\mathcal{R}_\text{su}$ is maximum for $X_- \approx 1.17$. It seems that this could be defined as a critical speed the swimmers cannot exceed. In all our observations, $\overline{u}<1.15$.

\subsection{Hill's heuristic law and propulsion time}
\label{HillsProp}
It is well known that the force decreases with the velocity (Hill's heuristic law). In this appendix, we discuss briefly this problem in our discussion of the propulsion time. We assume that the athletes can control their force of propulsion and that they can chose a bounded thrust $T_a \in\left[0,T_a^*\right]$. This thrust comes from the arms (in the present discussion the legs are tied), which are controlled by the swimmers muscles. Then the force used to activate the arms should be bounded by a parameter, which follows a Hill's law, and dynamic equation of the hand becomes:
\begin{align}
\alpha_h u_{h/w}^2 = \min\left\{ \varphi, \Phi^*\left(1-\dfrac{u_{h/w}}{u_h^*} \right)\right\},
\label{hillshand}
\end{align}
where $\varphi \in\left[0,1\right]$ is the control on the force, $\Phi^*$ and $u_h^*$ denote the maximum force that can be used to activate the arm and the maximum velocity at which this arm can be moved without resistance, respectively. Note that if eq.\ref{hillshand} is bounded by the Hill's law then $\Phi^* \left(1-{u_{h/w,\text{max}}}/{u_h^*} \right)=1$ at the maximum hand velocity in the water $u_{h/w,\text{max}}$, since we non-dimensionalize the problem here. And then we go back to what was done earlier. 

Further note that $u_h^*$ can also provide information on the air recovery time and thus the maximum index of coordination the swimmer can achieve $\IdC_{\max}$. Let's assume for now that the swimmer is pushing with the maximum thrust at $\tau_0$ and wants to swim at the maximum index of coordination. Then they have to bring the arm back to the front as fast as they can. They will reach $\approx u_h^*$ in the air recovery phase. 

\subsection{Impact of the legs kicking on the coordination}
\label{LegsCoord}
In all the paper the legs were tied. In previous works done by Chollet \textit{et al.} \cite{chollet2000} and Seifert \textit{et al.}\cite{seifert2007}, the swimmers were allowed to used their legs. The velocities the swimmers could reach were larger by 20\% compared to the present one for similar level swimmers. We look in this appendix at other data taken on similar level swimmers with the legs free to kick. We did not measure the resistance coefficient for these swimmers with the MAD-system. They only performed the coordination test.
 \begin{figure}[h!]
\centering
\includegraphics[width=.8\linewidth]{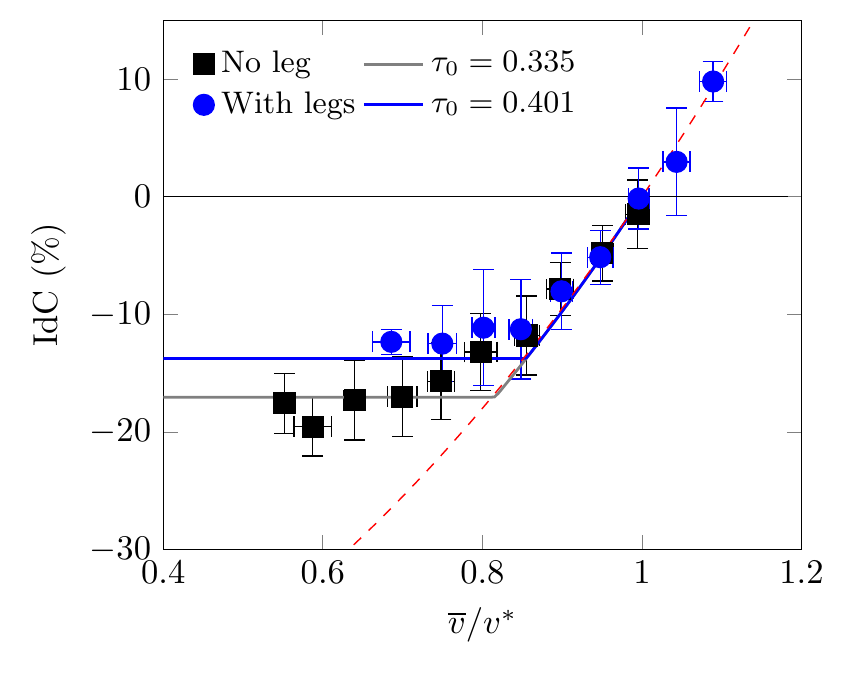}
\caption{Arm coordination with the mean non-dimensional velocity for swimmers with and without legs. The dashed line corresponds to the maximum force model. The solid lines show the optimal coordination with $\tau_0=0.335$ and $\tau_{0,\dagger}=0.401$ in gray and blue, respectively. In both cases, we used $\epsilon=0.035$.
}\label{IdCVNoDimwithLegs}
\end{figure}
  
We applied the same analysis as done in the present paper to the data of velocity and coordination. The results are displayed in figure \ref{IdCVNoDimwithLegs} and compared to the swimmers with no legs of the present paper. We observe a similar trend. The swimmers select a constant index of coordination at low velocity and then follow the maximum force model. The characteristic velocity $v^*$ increased from 1.5 m/s without leg to 1.8 m/s with legs in average.

At low velocity, the swimmers usually perform one kick per arm stroke and at higher velocity up to three \cite{millet2002coordination}. We will assume it is linked to the force choice of the arms and also in phase. Then the effect of the legs can be whether a reduction of the effective drag $k_b$ or an increase of the propulsion force $T_a$. In all cases, it will lead to an increase of the characteristic velocity $v^*$ and the coordination is certainly affected by this.

In this appendix, we will assume it is only the thrust that is increased and therefore the drag is of the same magnitude ($k_b = 30$ kg/m). Then if the physical propulsion time is the same, the non-dimensional one should be rescaled $\tau_{0,\dagger} = \tau_0 v^*_{\dagger}/v^*$, where the $\dagger$ indicates the quantity with the legs free to kick. Using for $\tau_0$ without leg the value found previously, we get $\tau_{0,\dagger}  = 0.401$ with legs. Our model then will predict for a similar $\epsilon$ an increase of the index of coordination from -17\% to -12\%. This is in good agreement with the observations.

\subsection{Coordination and swimming distances}
\label{coordinationLengthrace}
In the paper we describe the intra-cycle variations dynamic. We optimized the choice of coordination such that the swimmers minimize their energy consumption. To have an order of magnitude of the distance reached by the swimmers with this technique, an energy equation is necessary:
\begin{align}
\dfrac{dE}{dt} = \sigma- \dfrac{1}{\eta}T_a\left(t\right)v\left(t\right),
\end{align}
where $E$ denotes the energy reserves of the swimmer, $\sigma$ is the maximum rate at which the oxygen is supplied to the muscles (it is equivalent to the VO$_2$ max) and $\eta$ a conversion efficiency of the chemical energy to propulsive energy (which we will assume to be a constant). 

We integrates this equation on the total duration of the race $t_r = L_r/\overline{v}$:
\begin{align}
E_r-E_0  \approx \sigma t_r -  \frac{1}{\eta}\frac{t_r}{T} \int_0^T T_a\left(t\right)v\left(t\right)\,\mathrm{d}t ,
\label{energyeq}
\end{align}
where $E_0$ is the anaerobic reserve at the beginning of the race and $E_r\geq 0$ the left energy at the end of the race. The approximation comes from the number of cycle the swimmer used $\approx t_r/T$ on the second term of the right-hand-side and the fact that we neglect the turns and the start for simplification. Behncke \& Brosowski \cite{behncke1987optimization} discuss a way to take these parts of the race into account. To keep the discussion simple and analytical, we consider the case of an infinite pool in the present discussion. The second term in the right hand side of eq.\ref{energyeq} is the one we minimised in the previous sections for a given mean velocity $\overline{v} = \overline{u} v^*$. 
Using the opposition mode as a reference eq.\ref{energyeq} becomes:
\begin{align}
E_r-E_0  \approx \sigma t_r - \dfrac{1}{\eta} t_r \mathcal{R}\left(\overline{u} \right) k_b \overline{v}^3,
\label{energyeq2}
\end{align}
where $\mathcal{R}$ is the economy  at $\overline{v} = \overline{u} v^*$ and will depend on the coordination. 

For short races (pure sprints, $E_r>0$), the swimmer does not have to worry about the economy and should select the highest possible velocity they can achieve. A superposition mode is expected with the highest possible index of coordination, as this is how they can achieve the highest velocity. They can maintain this technique as long as $E_r>0$. We define the aerobic velocity: \begin{align}
v_{\sigma} = \left(\dfrac{\eta \sigma}{k_b}\right)^{\sfrac{1}{3}},
\end{align}
the characteristic distance:
\begin{align}
L_0 = \dfrac{\eta^{\sfrac{1}{3}} E_0}{\sigma^{\sfrac{2}{3}}k_b^{\sfrac{1}{3}}}.
\end{align}
and:
\begin{align}
\beta  = \dfrac{v_\sigma}{v^*}.
\end{align}
It then comes that the sprint technique will last as long as $L_r<L_s$, where:
\begin{align}
L_s =  \dfrac{L_0}{\mathcal{R}^{\max}\left(u_{\max}/\beta\right)^2 - \beta/u_{\max}},
\end{align}
where $\mathcal{R}^{\max}$ is the economy at the maximum index of coordination and force, and $u_{\max}$ the maximum non-dimensional speed. At worst, it will corresponds to the maximum for the superposition mode (see appendix \ref{extensionSuperposition}) which occurs for $u_{\max} \approx 1.17$. We will use this extreme in the applications. Obviously the numerator should be positive in this expression. Otherwise, it would mean that the swimmers could sprint as long as they want. This will give an upper bound to the possible values of $\eta$.

For longer races, $E_r=0$ and the swimmer should minimize their energy consumption. They should then choose the coordination that enables to minimize the economy at the targeted mean velocity. 
Eq.\ref{energyeq2} becomes:
\begin{align}
0-E_0  \approx \sigma t_r - \dfrac{1}{\eta} t_r \mathcal{R}^{op}\left(\overline{u} \right) k_b \overline{v}^3,
\label{energyeq3}
\end{align}
where $\mathcal{R}^{op}$ is the optimal economy  at $\overline{v} = \overline{u} v^*$\footnote{for a velocity below the critical velocity $v_c = u_c v^*$, it is $\mathcal{R}_\text{cu}^{\min}$ discussed in the main text.}. 

The link between the length of the race and the mean velocity is then:
\begin{align}
\dfrac{L_0}{L_r} = \mathcal{R}^{op}\left(\overline{u}\right) \left(\dfrac{\overline{u}}{\beta}\right)^2  - \dfrac{\beta}{\overline{u}}.
\label{Lrv0Relation}
\end{align}

Eq.\ref{Lrv0Relation} gives the relationship between the mean velocity and the length of the race depending on the swimmer characteristics (VO$_2$ max, anaerobic velocity, gliding efficiency and propulsion time). The transition from the long distance race strategy with constant index of coordination and the shorter race with maximum force occurs at the $\overline{v} = u_c v^*$. Injecting the results of the paper for $\mathcal{R}_\text{cu}^{\min}$ and $u_c$ in eq.\ref{Lrv0Relation}, it comes:
\begin{align}
L_c = \dfrac{L_0}{ \left(u_c/\beta\right)^2 \mathcal{R}^{\min}_\text{cu} - \beta/u_c}.
\label{LcEq}
\end{align}
This transition will occur only if the denominator in eq.\ref{LcEq} is positive. Otherwise, the swimmer will stay in the maximum force model for all the distances. Note that we do not take into account fatigue here. 

To have an order of magnitude of these different lengths, we consider typical values of the different parameters. We will consider a swimmer with the legs free to kick (and therefore $\tau_0=0.4$ and $\epsilon = 0.035$, see appendix \ref{LegsCoord} for the origin of the value of $\tau_0$). We consider an athlete of 
$m_0=80$ kg, with a typical drag $k_b = 30$ kg/m, a characteristic velocity $v^* = 1.8$ m/s, $E_0 = 152$ kJ, $\sigma = 2.08$ kJ/s and $\eta=0.04$ \cite{keller1974optimal,keller1973ia,behncke1987optimization,di1974energetics}. Note that we choose $\eta$ to be constant. di Pramparo \textit{et al.} \cite{di1974energetics} evaluated the efficiency in sub-maximal exercise to be in between 2.6 and 5.2\%. This value seems therefore reasonable. Note that to be consistent, here the efficiency cannot be larger than 13.5 \% because then the $L_s$ would not be defined. The choice of the efficiency greatly influences the results. With the present values, we get $L_0=103$ m, $v_\sigma = 1.40$ m/s and $\beta = 0.78$. It then comes that $L_s = 64.7$ m and $L_c = 301$ m. A pure sprint is limited to the 50 m race and the 100 m race swimmers are expected to manage their energy on the length of the race. The swimmer will switch from a maximum force technique to a constant index of coordination race in between the 200m and the 400m races. This is consistent with the observations of Craig \& Pendergast \cite{craig1979relationships} and previous observations of change in coordination patterns with pace made by Chollet \textit{et al.} \cite{chollet2000} and Seifert \textit{et al.}\cite{seifert2007}.

\end{document}